\documentclass[12pt]{article}
\usepackage{pdproc,epsfig} 

\begin{document}
\newcommand{\minerva}{\mbox{\hbox{MINER}$\nu$\hbox{A}}}
\newcommand{\nova}{\mbox{\hbox{NO}$\nu$\hbox{A}}}

\title{ Neutrino Scattering Uncertainties and \\ 
their Role in Long Baseline Oscillation Experiments }

D.~A.~Harris$^4$, 
G.~Blazey$^{10}$, 
A.~Bodek$^{13}$, 
D. Boehnlein$^4$, 
S.~Boyd$^{12}$, 
W.K.~Brooks$^{11}$, 
A.~Bruell$^{11}$, 
H.~Budd$^{13}$, 
R.~Burnstein$^6$, 
D.~Casper$^2$, 
A.~Chakravorty$^{6,7}$,
M.E.~Christy$^5$, 
J.~Chvojka$^{13}$, 
M.A.C.~Cummings$^{10}$, 
P.~deBarbaro$^{13}$, 
D.~Drakoulakos$^1$, 
J.~Dunmore$^2$, 
R.~Ent$^{11}$, 
H.~Gallagher$^{15}$, 
D.~Gaskell$^{11}$, 
R.~Gilman$^{14}$, 
C.~Glashausser$^{14}$, 
W.~Hinton$^5$,
X.~Jiang$^{14}$, 
T.~Kafka$^{15}$, 
O.~Kamaev$^6$, 
C.E.~Keppel$^{5,11}$, 
M.~Kostin$^4$, 
S.~Kulagin$^8$,
G.~Kumbartzki$^{14}$,
S.~Manly$^{13}$, 
W.A.~Mann$^{15}$,
K.~McFarland$^{13}$,
W.~Melnitchouk$^{11}$, 
J.G.~Morf\'{\i}n$^4$,
D.~Naples$^{12}$, 
J.K.~Nelson$^{16}$, 
G.~Niculescu$^9$,
I.~Niculescu$^9$, 
W.~Oliver$^{15}$, 
V.~Paolone$^{12}$,
E.~Paschos$^3$,
A.~Pla-Dalmau$^4$, 
R.~Ransome$^{14}$,
C.~Regis$^2$, 
P.~Rubinov$^4$, 
V.~Rykalin$^{10}$,
W.~Sakumoto$^{13}$,
P.~Shanahan$^4$, 
N.~Solomey$^6$,
P.~Spentzouris$^4$,
P.~Stamoulis$^1$, 
G.~Tzanakos$^1$, 
S.A.~Wood$^{11}$,
F.X.~Yumiceva$^{16}$,
B.~Ziemer$^2$,
M.~Zois$^1$ 

\centerline{\it $^1$University of Athens; Athens, Greece}
\centerline{\it $^2$University of California, Irvine; Irvine, California, USA}
\centerline{\it $^3$University of Dortmund, Dortmund, Germany}
\centerline{\it $^4$Fermi National Accelerator Laboratory; Batavia, Illinois, USA}
\centerline{\it $^5$Hampton University; Hampton, Virginia, USA}
\centerline{\it $^6$Illinois Institute of Technology; Chicago, Illinois, USA}
\centerline{\it $^7$Saint Xavier University; Chicago, Illinois, USA}
\centerline{\it $^8$Institute for Nuclear Research, Moscow, Russia}
\centerline{\it $^9$James Madison University, Harrisonburg, Virginia, USA}
\centerline{\it $^{10}$Northern Illinois University; DeKalb, Illinois, USA}
\centerline{\it $^{11}$Thomas Jefferson National Accelerator Facility; 
Newport News, Virginia, USA}
\centerline{\it $^{12}$University of Pittsburgh; Pittsburgh, Pennsylvania, USA}
\centerline{\it $^{13}$University of Rochester; Rochester, New York, USA}
\centerline{\it $^{14}$Rutgers, The State University of New Jersey; 
Piscataway, New Jersey. USA}
\centerline{\it $^{15}$Tufts University; Boston, Massachusetts, USA}
\centerline{\it $^{16}$William and Mary College, Williamsburg, Virginia, USA} 

\noindent

\abstract{ The field of oscillation physics is about to make an 
enormous leap forward in statistical precision: first through the 
MINOS experiment in the coming year, and later through the 
\nova\ and T2K experiments.  Because of the relatively poor understanding of
neutrino interactions in the energy ranges of these experiments, 
there are systematics
that can arise in interpreting far detector data that can be as large
as or even larger than the expected statistical uncertainties.  
We describe how these systematic errors arise, and how specific 
measurements in a dedicated neutrino scattering experiment 
like \minerva\ can reduce the cross section systematic errors to well 
below the statistical errors.  } 

\section{Introduction}
 
     Over the past 5 years the 
field of neutrino oscillations has moved from seeing decade-old anomalies in 
cosmic ray~\cite{atm} and solar~\cite{sol} neutrino data to 
cross checks of these anomalies (SNO data~\cite{SNO} 
and angular distributions in 
atmospheric neutrino data~\cite{skatmall}) 
and most recently to terrestrial confirmations of the 
oscillation hypothesis (Kamland~\cite{KAMLAND} and K2K~\cite{K2K}).  
The next steps in this field are to 
1) move to the precision realm of measurements of the mass splittings and the 
mixing angles that have been observed, and 2) to see if
any more off-diagonal elements in the neutrino mixing matrix are non-zero.  
 
New extremely intense beamlines are being built or planned that will
greatly increase the statistical reach and ultimate precision on
oscillation parameters.  However, with such large improvements in the
statistical accuracy come new concerns about systematic uncertainties
that have until now been negligible.  In particular, uncertainties
in neutrino cross sections and nuclear effects can produce systematic
uncertainties in the extraction of mixing parameters.  Although near
detectors are a critical part of precision long-baseline oscillation
measurements, they are not often well-suited to make all the needed cross
section measurements, due to the fact that they tend to be very
similar to the massive far detectors.  Furthermore, a near detector
can at best be a constraint on the product of the near flux, cross
section and detection efficiency.  Uncertainties on all of these
quantities must be incorporated in ultimate near detector analyses.
The studies described in this document do not address these other
uncertainties, but when taken into account clearly worsen the prediction
from the near detector data beyond what is described here.  
 
This article is divided into two sections.  The first section addresses
the kinds of uncertainties that are most relevant for $\nu_\mu$
disappearance experiments, whose aim is to precisely measure the mass
splitting $\Delta m^2_{23}$, and the mixing angle which has already
been determined to be large, $\theta_{23}$.  In order to achieve these
goals the experiments must measure oscillation probabilities as a
function of neutrino energy.  Two important concerns here are
uncertainties in charged current non-quasi-elastic processes, and the
scale of nuclear effects.  Both non-quasi-elastic channels and the 
nuclear environment alter the
relationship between the measured and true neutrino energy.
The second section addresses
experiments searching for $\nu_e$ appearance, which if seen would
indicate a non-zero value of $\theta_{13}$.  Because the size of the
signal is unknown, the final event sample may be dominated by both
signal (charged current) cross sections, or by background (neutral and
charged current) processes.  Either way, the experiments of the past
are not precise enough to provide accurate predictions for the
far detector event samples.  

After discussing the ways neutrino
interaction uncertainties apply to each of these measurements, a
description is given of the kind of neutrino scattering 
measurements that are needed.  As an example we give the expected
precision of the \minerva\ experiment, which has been 
proposed to run parasitically in the NuMI beamline \cite{minervaprop}.  

\section{$\nu_\mu$ Disappearance} 

In order to precisely measure the mass splitting between two
eigenstates one must measure the oscillation probability as a function
of neutrino energy ($E_\nu$) divided by baseline ($L$).  The muon
neutrino disappearance probability (in the standard 3-generation
oscillation parameterization~\cite{3genparam}) is expressed as

\begin{equation} 
P(\nu_\mu \to \nu_\mu) = 
1 - \cos^4 \theta_{13}\sin^2 2\theta_{23}\sin^2 \left( \frac{1.27\Delta m_{23}^2(eV^2) 
L(km)}{E_\nu(GeV)} \right) -...
\end{equation} 
where the additional terms are $\cal{O}(\sin^2 2\theta_{13})$ or smaller.  
Currently $\Delta m^2_{23}$ is known to within a factor of two and
$\cos^4\theta_{13}\sin^2 2\theta_{23}$ has been shown to be above 0.9, 
at the 90\% confidence level limit~\cite{valle}.  Since 
$\sin^2 2\theta_{13}$ has been limited to below 0.1 by the CHOOZ reactor
experiment\cite{chooz}, this means that $\sin^2 2\theta_{23}$ itself is very
close to 1.  The fact that $\theta_{23}$ is close to $45^\circ$ has
been cited as a hint of the underlying symmetry that generates
neutrino mass and mixing.  Precise measurements of this angle are
important because the level at which the mixing deviates from maximal
may again give hints to possible mechanisms for the breaking of that
symmetry~\cite{symmetry}.  
Furthermore, more precise measurements $\Delta m_{23}^2$
are required to extract mixing angles from eventual $\nu_e$ appearance
measurements.

The challenge of measuring $\Delta m^2_{23}$ lies in knowing the true
neutrino energy in both near and far detectors.  Even if the two
detectors have an identical design, any uncertainty in the ``neutrino
energy scale'' of the signal events translates directly into an
uncertainty in the extracted value of $\Delta m_{23}^2$.
There are two different ways of measuring neutrino 
energies: either kinematic or calorimetric reconstruction.  
We discuss both techniques
here, and then explain how uncertainties 
in neutrino interactions translate into energy scale 
uncertainties and ultimately $\Delta m^2_{23}$ uncertainties.  

The first experiment to provide a precision measurement of $\Delta m^2_{23}$ 
will be the MINOS experiment \cite{minos}, which will start 
taking data at the beginning of 
2005.  The MINOS experiment will use both far and near detectors, 
which consist of 
magnetized steel-scintillator calorimeters with a longitudintal 
steel segmentation of 2.54~cm.  The transverse 
segmentation of the 1~cm thick scintillator planes is 4~cm.   
The MINOS experiment, with a baseline of 735~km, 
will use the NuMI beamline located at Fermilab, which can provide a variety 
of broad band neutrino spectra.  
In its lowest energy configuration, which is where 
MINOS expects to do most of its running, the peak neutrino 
energy in the $\nu_\mu$ event spectrum is about 3.5~GeV.

The second experiment to use a calorimetric detector 
and improve the measurement 
of $\Delta m_{23}^2$ is \nova\ . Because \nova\ is 
optimized for $\nu_e$ appearance
rather than $\nu_\mu$ disappearance it will use near and 
far calorimeters made of scintillator
planes interspersed with either particle board or with 
other scintillator planes.  
The longitudinal segmentation is expected to be about a third to a sixth of a 
radiation length, and the transverse 
segmentation of the scintillator will be about 4~cm\cite{novaprop}.   
\nova\ will also use the NuMI beamline, but will place its 
detectors between 12 and 14~mrad off the 
beamline axis, to get a narrow band neutrino spectrum.  
\nova\, with a baseline of 810~km, will run with a peak 
neutrino energy of about 2~GeV.  

Finally, the T2K experiment will use the Super-Kamiokande 
water Cerenkov detector for its far detector, 
and focus on single-ring muon-like 
events, for which the neutrino energy reconstruction is kinematic.  T2K will 
use a narrow band  neutrino beam from J-PARC in Tokai, whose peak 
is close to 700~MeV and which originates some 295~km from 
the far detector \cite{t2kbeam}. 
The near detector design has not been finalized, but at the time of 
this writing a water Cerenkov near detector 
is not forseen as part of the first phase of the experiment.  

\subsection{Kinematic Recontruction of Neutrino Energy} 

In kinematic reconstruction one 
assumes that the event is of a particular process (for example, quasi-elastic) 
and one calculates the energy assuming the kinematics of that reaction.  This is 
the technique that is used predominantly in water Cerenkov detectors, which 
operate best in regimes where the quasi-elastic process dominates the cross section.  
In the Super-Kamiokande detector, for example, the $\nu_\mu$ charged current  
signal sample consists of single ring muon-like events, which are then 
assumed to be quasi-elastic events.  The energy of the incoming neutrino can 
in that case be 
calculated using only the outgoing muon momentum ($p_\mu$) 
and direction ($\theta_\mu$), as follows:  

\begin{equation} 
E_\nu = \frac{m_N E\mu - m_\mu^2/2}{m_N - E_\mu + p_\mu \cos\theta_\mu} 
\end{equation} 

Since the absolute energy scale for muons can be known to 
better than 1\% through a variety of calibration techniques \cite{sknim}, 
and the ring-finding algorithms can measure ring 
directions extremely well, it seems plausible that the neutrino energy scale
would also be determined to better than 1\%.  However, not all events
that pass a ``single muon-like ring'' cut are quasi-elastic events.  There
are resonance and deep inelastic events where one or more 
pions have been absorbed in the nucleus, or which have one or more pions below
the Cerenkov threshold, and those events will have a reconstructed energy
which is well below the true neutrino energy, while still passing all cuts. 
Uncertainty in the ratio between quasi-elastic and resonance 
cross sections as a function of energy produces an uncertainty in the 
effective neutrino energy scale of the detector.  Furthermore, because the 
$\nu_\mu$ disappearance probability is large where T2K will run, the mix of 
quasi-elastic to non-quasi-elastic events will be very different from the 
mix one would expect if there
were no $\nu_\mu$ disappearance (which would also be the case for the 
mix at a near detector).  

\begin{figure}[tp]
\begin{minipage}{.5\textwidth}
\epsfxsize=\textwidth
\epsfbox[20 160 550 650]{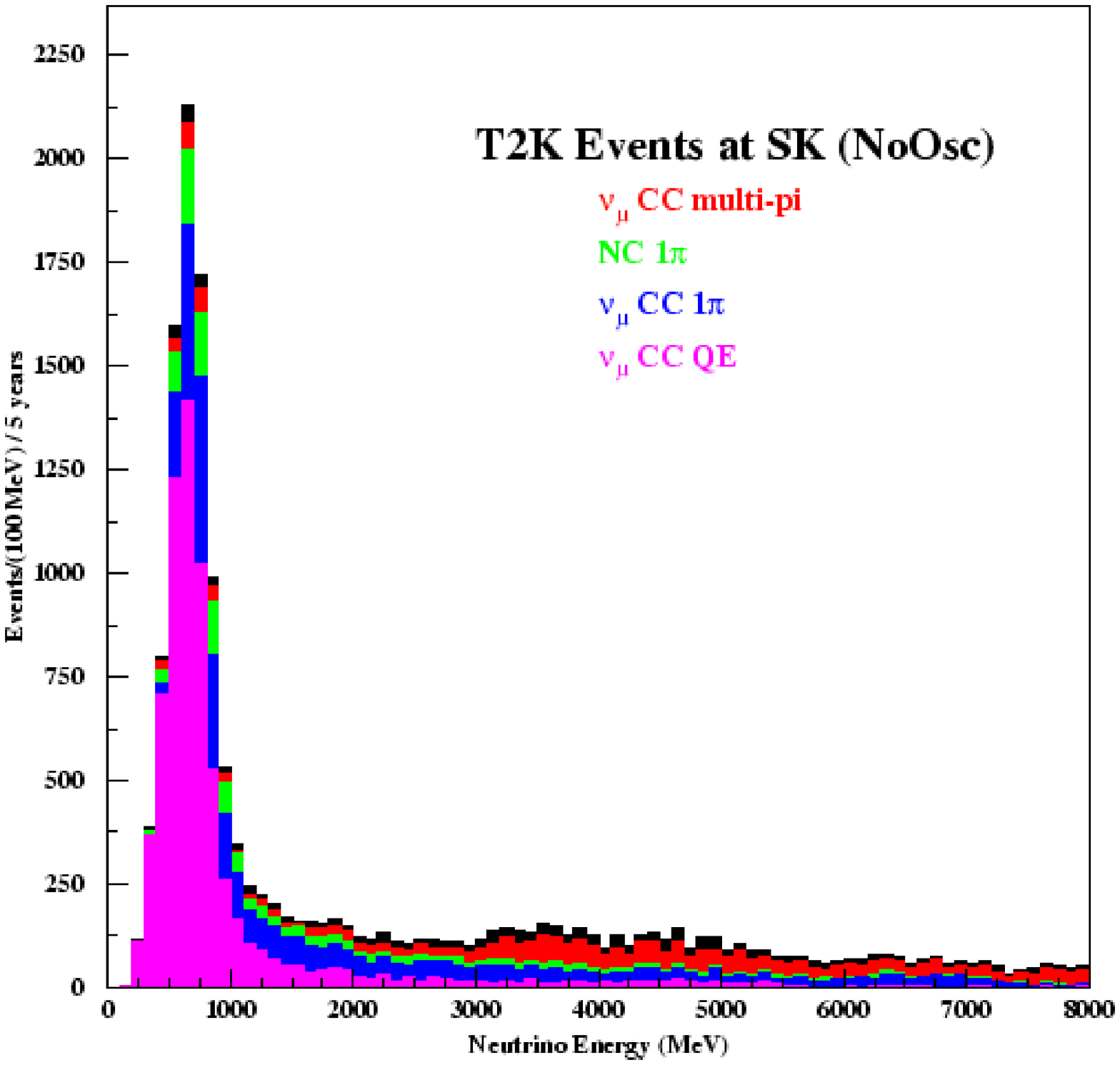}
\end{minipage} 
\begin{minipage}{.5\textwidth}
\epsfxsize=\textwidth
\epsfbox[20 160 530 650]{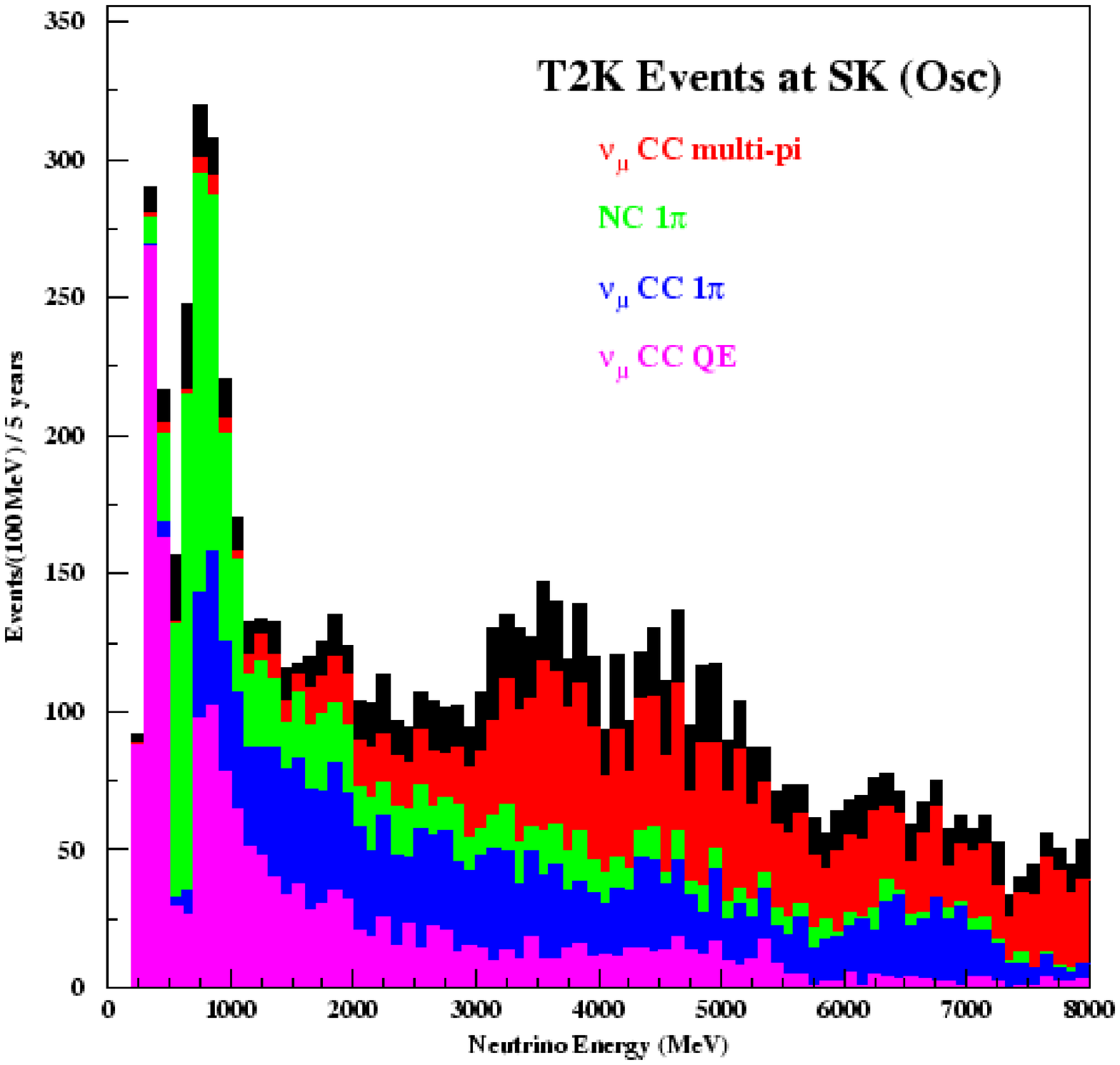}
\end{minipage} 
\caption{The neutrino energy distribution for 
events at T2K, broken up into various processes:  quasi-elastic,
single pion (Resonance), multi-pion (DIS), and neutral currents, for 
(left) no oscillations and (right) oscillations }
\label{fig:dave}
\end{figure}

To understand how different the mix of signal processes is, consider
the event spectra from the T2K beamline at the Super-Kamiokande
detector, with and without oscillations.  The NUANCE neutrino event 
generator \cite{nuance} was used with fluxes from the T2K beamline
simulation \cite{t2kbeam}.  
Figure \ref{fig:dave} shows all of the contributions
to the far detector event sample in the T2K experiment, without (left) 
and with (right) 
oscillations, after 5 years of running at the expected 
intensity.  Note that in the case of no oscillations the event 
sample is predominantly quasi-elastic, but with oscillations the quasi-elastic
contribution is much smaller and there are important contributions from resonant 
processes (single pion) and even deep inelastic scattering processes (multi-pi). 

\subsection{Current and Future Measurements of the Quasi-Elastic and Non-Quasi-Elastic 
Cross Sections} 
\label{section:xsecerr} 

\begin{figure}[tp]
\begin{minipage}{.625\textwidth}
\epsfxsize=\textwidth
\epsfbox{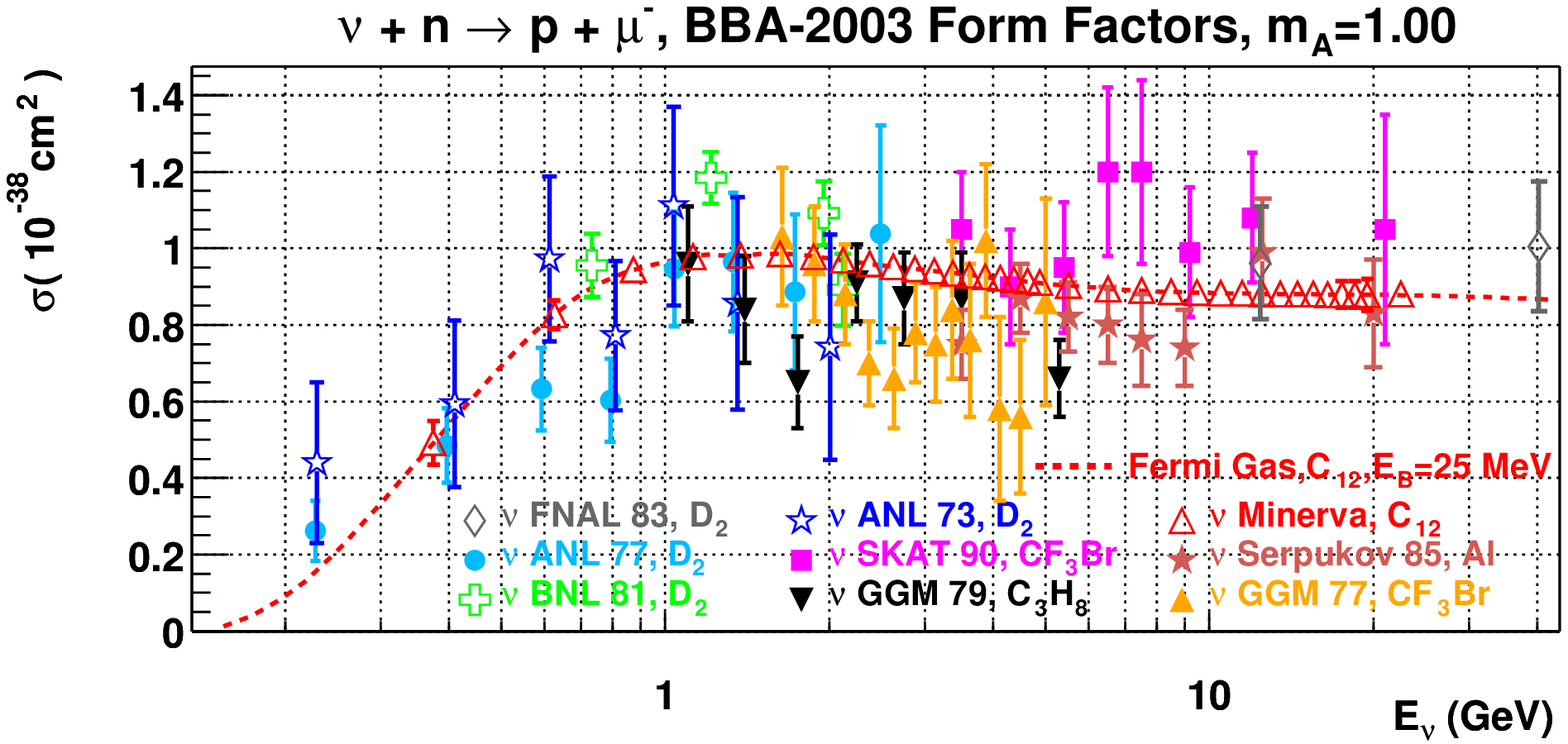} 
\end{minipage} 
\begin{minipage}{.375\textwidth}
\epsfxsize=\textwidth
\epsfbox{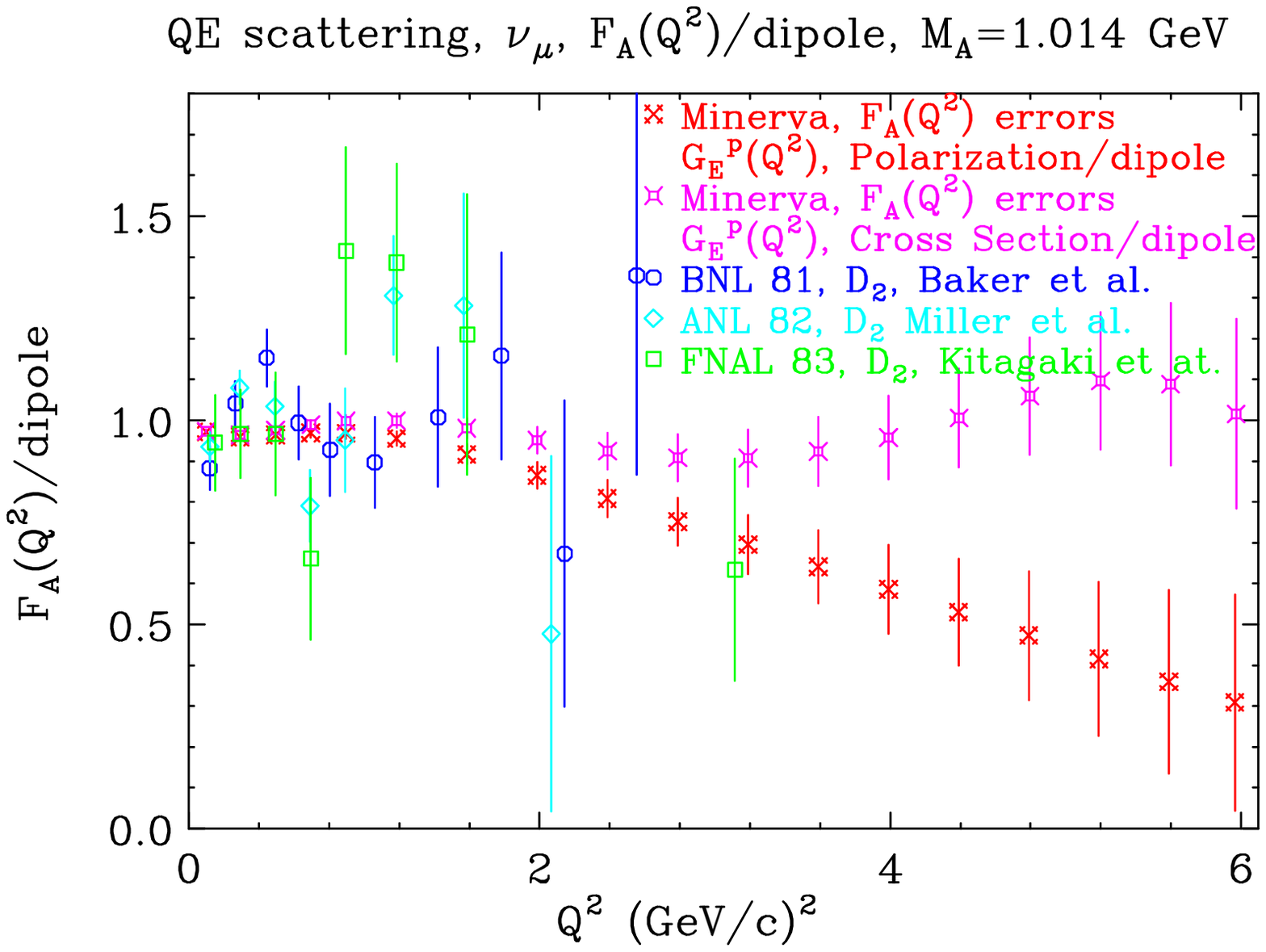} 
\end{minipage} 
\caption{Current and expected \minerva\ statistical 
sensitivity for quasi-elastic cross section (left) and form factor (right) 
measurements, for a 4 year parasitic MINOS run.
Left:  the open red triangles are in many energy bins larger than 
the statistical error expected in \minerva\, 
taking into account detector
acceptance and resolution.}
\label{fig:minqe}
\end{figure}

As is shown in figure~\ref{fig:minqe}, 
the quasi-elastic cross sections themselves are known to at best the 10\% level, and 
worse at energies of a few GeV \cite{xsecqe}.  
Current measurements of the charged current single pion and multi pion cross sections 
come from experiments done in the 80's \cite{ccsinglepiplus,ccmultipi}, 
and are known to at best the 20\% level~\cite{sam}.  However, 
some of these measurements have 
central values which differ by much more than the total error bars, and the 
cross sections were measured 
on a variety of neutrino targets. The K2K experiment
has a fine-grained near detector which can try to measure the non-quasi-elastic
to quasi-elastic ratio.  In reference \cite{K2K} this ratio was 
assigned an error of 20\% 
based on considering different cross section models which were all in agreement
with their near detector data.   
One can see that the statistical error for the final event sample will 
be well above 100 events in total, 
so future constraints of this ratio will be extremely important. 
 
What would best reduce this uncertainty for future experiments 
are precise measurements of both the differential 
single-pion and multipion charged current 
cross sections, as a function of neutrino energy.  Clearly because the 
event samples are so different between near and far detectors, and because
the water Cerenkov technology is not enough to constrain this ratio, 
additional measurements with fine-grained detectors are required.  Ideally, 
there would be measurements of exclusive non-quasi-elastic 
final states identified with a well-modeled efficiency relative 
to that of quasi-elastic events.  
Because the reconstructed energy for these events is lower than the true
neutrino energy, it is important to measure 
the charged current single and multi-pion (resonance) cross sections 
both at and above the T2K neutrino energy.

By identifying both the outgoing muon and 
proton in a quasi-elastic event, and by requiring there to be no other 
outgoing track, a fine-grained detector such as the one proposed by 
\minerva\ can cleanly separate quasielastic events in a broad 
energy range, and the expected purity is  
above 70\% \cite{minervaprop}.  
In 4 years of parasitic NuMI running 
\minerva\ hopes to collect about $10^5$ Quasielastic events per ton, 
and the expected 
statistical error on the cross section precision as a function of energy  
(after taking into account detector acceptance, backgrounds, and resolution) 
is shown in figure \ref{fig:minqe} (left).  Figure \ref{fig:minqe} (right) 
shows how \minerva\ would also have adequate statistics and resolution 
to discriminate between two different models for the $Q^2$ dependence of 
the quasi-elastic form factor, which again will have relevance for 
the quasi-elastic to non-quasielastic ratio.  
The systematic error in the energy dependence would most likely be
dominated by the flux uncertainty, coming from the MIPP data on hadron 
production \cite{mipp}, but is expected to be at the 5\% level at low 
energies.  

\subsection{Calorimetric Energy Reconstruction} 

At neutrino energies higher than 1~GeV, calorimetric energy
reconstruction is a much more useful technique than kinematic
reconstruction.  In a calorimetric device the reconstructed or visible
neutrino energy is simply the sum of all the secondary particles'
energies that are visible in the event.  For a $\nu_\mu$ charged current event,
the muon energy can be determined by first measuring its momentum
using either range or curvature (if the calorimeter is magnetized),
and then the remaining signal in the event is summed to be the hadron
energy.  Because most calorimeters have a much lower pion threshold
than Cerenkov detectors, much more of the total kinetic energy is
visible for multi-pion events, which dominate the cross section above
a few GeV.  As a result, the neutrino energy reconstruction is not as
biased for non-quasi-elastic events as it is for water Cerenkov
detectors.

     For the MINOS detector, the absolute energy scale of the 
muons is set by knowing the thickness of the steel plates and 
by understanding the process of muon energy loss.   
The thickness of each of the plates has been measured to better than 0.1\% and 
they vary with an RMS of 0.4\%~\cite{daveb}.  A muon test beam was used 
at CERN where a 2\% absolute scale calibration was  
achieved \cite{muontheses}.  
The hadronic and electromagnetic energy scales have been calibrated
using test beams on a prototype detector at CERN, and have been
measured relative to the muon scale to better than 5\%
\cite{minoshadrons,minoselectrons}.  However, one must translate from
the response from pions and muons to that of interacted neutrinos.  

At neutrino energies of a few GeV and below, there are three effects that
become significant in the translation between between visible energy
and neutrino energy.  Uncertainties in these effects must be
understood and included in any precise measurement of $\Delta
m_{23}^2$.  One effect, which is independent of the target nucleus, is
the fact that of the rest masses the secondary charged pions become
important.  Since MINOS cannot measure the multiplicity of final state
particles, a multiplicity distribution 
as a function of hadron energy must be assumed.  The second and
third effects are due to the fact that secondary particles can either
scatter in the nucleus or be completely absorbed.  All three of these
effects result in a reduction in the visible hadron energy in an
event, which therefore results in a lower reconstructed neutrino
energy.  As is described in reference \cite{Paschos}, the size of
these effects can be quite large as the parent neutrino energy
decreases, since there is a peak in the pion absorption cross section
for pions at several hundred MeV~\cite{ashery}.

In order to evaluate the extent to which nuclear effects will alter a
$\Delta m_{23}^2$ measurement in a MINOS-like detector, a crude
detector simulation combined with the NEUGEN event
generator~\cite{neugen} and NuMI fluxes at 735~km~\cite{markflux} was used.
In this simulation the visible 
energy is defined simply as the sum of the kinetic energies of
all the charged final state particles, plus the total energy for the
neutral pions, and photons, since it is assumed they deposit all their
energy in the form of electromagnetic showers.
 
\begin{figure}[tp]
\epsfxsize=\textwidth
\epsfbox{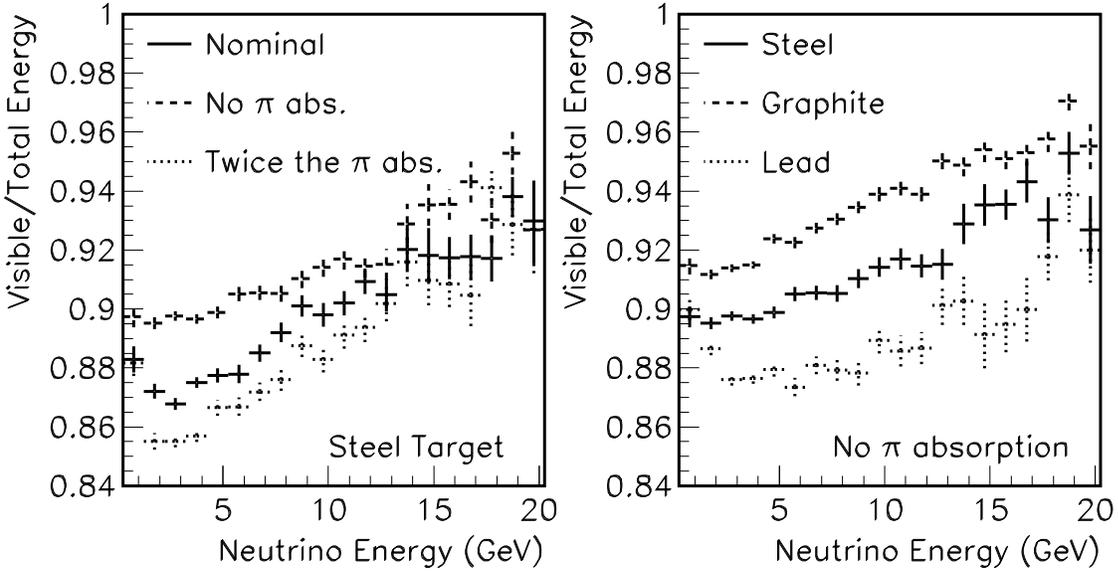}
\caption{Ratio of visible (reconstructed) 
to true neutrino energy for several different 
models of nuclear effects.  The left plot shows the ratio for steel (solid) 
with the nominal pion absorption, as well as the same ratio for the pion
absorption turned off or doubled above what is expected.  
The right plot shows the differences the ratio for three 
different target nuclei, where the pion absorption effects are turned off to 
isolate the effects of pion rescattering.}
\label{fig:oscplot_a}
\end{figure}

Figure \ref{fig:oscplot_a} shows the changes in the ratio of visible
to total neutrino energy for changes in absorption and scattering
separately.  For the plot on the left the target is assumed to be
steel, and the parameter in the event generator that describes pion
absorption is set to zero or doubled.  For the plot on the right all
pion absorption is turned off, and the differences that remain are due
to the rescattering effects between steel, carbon, and lead.  Because
the $\nu_\mu$ disappearance probability is expected to be large, the
far and near detector energy spectra will be very different, and therefore
these effects will only partially cancel between the near and far
detector.  The extent to which they do not cancel results in a
systematic error on $\Delta m^2_{23}$.

\begin{figure}[tp]
\begin{center} 
\epsfxsize=.5\textwidth
\epsfbox{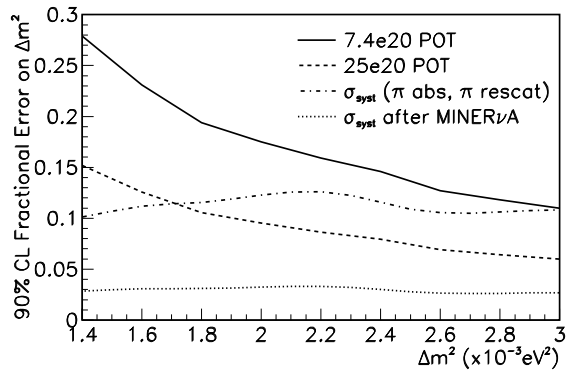}
\end{center} 
\caption{Fractional size of the 90\% confidence level region at
$\sin^2 2\theta_{23}=1$ from statistics for the MINOS experiment.
Also shown are possible systematic uncertainties due to 
uncertainties in nuclear effects:  
the dot-dashed line are those effects described in the text, and 
the dotted line assumes uncertainties after dedicated nuclear effect 
measurements where pion rescattering and absorption are measured 
on the target nucleus (steel).  Detector acceptance is 
modelled by requiring muons to be above 0.5~GeV. Also shown are the 
statistical errors for two different integrated proton intensities.}
\label{fig:oscplot_d}
\end{figure}

If we take the two differences described above as the uncertainties in 
pion absorption and rescattering, we can determine how this would 
compare to the MINOS statistical error.  In a more complete analysis, the 
detector acceptance must also be taken into 
account.  The most important cut that will 
reduce the size of nuclear effects comes from requiring
the muon to take up a minimum energy in the event.  
The smaller the neutrino energy
that comes from the hadron contribution, the smaller the 
changes which the nuclear effect uncertainties 
will bring to the total neutrino energy measurement.  
However, by requiring the muon 
to take up most of the neutrino energy, 
one will be losing precious far detector 
statistics.  In the evaluation of the systematic errors shown here, 
a minimum muon energy 
cut of 0.5~GeV was made to try to take into 
account the acceptance in a real analysis.  
If the uncertainties on nuclear effects are assigned 
to be the differences shown in figure \ref{fig:oscplot_a}, 
then with a 0.5~GeV muon momentum cut they induce an error in $\Delta m^2_{23}$
that is only slightly smaller than the statistical 
error expected by MINOS for $7.6\times 10^{20}$ protons on target (POT), 
as shown in figure \ref{fig:oscplot_d}.   

\subsection{Current and Future Measurements of Nuclear Effects in Neutrino Scattering} 
 
Evaluating the appropriate uncertainty in the size of nuclear effects
in neutrino scattering is not trivial, because the only data on these
effects in heavy nuclei come from charged lepton
scattering~\cite{emc}, and one has to use theoretical models to
translate the effects from the charged leptons to the neutral leptons.
The only neutrino data measuring nuclear effects with neutrinos comes
from pion rescattering measurements on Ne and $D_2$\cite{intranuke}.

In order to make a precise measurement of nuclear effects in neutrino
scattering one should measure
interactions on several different target nuclei simultaneously, 
where one of the
nuclei is the same as the far detector, and the other targets span a broad
range of atomic number.  A detector which can precisely identify the
target nucleus event-by-event is critical.  In this way the nuclear
effects and their energy dependence can be measured at least in
charged current interactions, and given a detector with good enough
$x$ and $Q^2$ resolution, these kinematic dependences can also be
measured.

The \minerva\ experiment has proposed a fine-grained detector 
which would measure neutrino interactions on steel, carbon,
and lead.  By running
parasitically in the NuMI beamline for four years, the experiment
would be able to collect about 940~k events on iron and lead, and 2.8M
events on carbon within the fiducial volume of the
scintillator\cite{minervaprop}.  This enormous improvement in both 
statistics and range of 
target nuclei would change our level of understanding of nuclear
effects in a fundamental way, and give real constraints on neutrino
interaction models.  The uncertainties in $\Delta m_{23}^2$ effects
with this new data in hand would be small compared to the statistical 
error, even for higher levels of integrated protons on target, as is shown in 
figure \ref{fig:oscplot_d}.
  
\section{$\nu_e$ Appearance} 

The goal of the next generation of neutrino
oscillation experiments is to determine whether or
not the last unmeasured neutrino mixing matrix element, (called
$|U_{e3}|$ or $\sin\theta_{13}$) is non-zero.  If $\theta_{13}$ is in
fact non-zero then there is a chance that future experiments can
search for CP violation in the lepton sector.  If it is non-zero then the 
possibility of measuring the neutrino mass hierarchy also arises.  For
T2K and \nova\, probing this matrix element is
done by measuring the $\nu_\mu \to \nu_e$ oscillation probability at a
``frequency'' corresponding to $\Delta m^2_{23}$.  The oscillation
probability for $\nu_\mu \to \nu_e$ in vacuum can be expressed as
~\cite{3genparam}
\begin{equation} 
P(\nu_\mu \to \nu_e) = \sin^2 \theta_{23}\sin^2 2\theta_{13}\sin^2 \left( \frac{1.27\Delta m_{23}^2(eV^2) L(km)}{E_\nu(GeV)} \right) + ...
\end{equation} 
where the additional terms not shown are due to effects from the
non-zero solar mass splitting, $\Delta m^2_{12}$.  
 
Looking for $\nu_e$ appearance in a $\nu_\mu$ beam is quite
challenging for several reasons.  According to the CHOOZ limit from
reactor neutrinos on $\sin^2 2\theta_{13}$\cite{chooz} the appearance
probability must be less than about 5\% at the 90\% confidence level.
Also, there is an intrinsic $\nu_e$ component that can be as large
as a few per cent.  Finally, neutral current or
high-$y$ charged current $\nu_\mu$ interactions can produce energetic
neutral pions, which can in turn produce electromagnetic showers that
fake a $\nu_e$ charged current event.
 
The T2K and \nova\ experiments will reduce these backgrounds
significantly below that of the current generation of long baseline
experiments by using detectors optimized for electron appearance, and
by placing those detectors off the beamline axis.  Because of the two
body decay of the charged pion, the energy spectra at small angles 
with respect to the beamline axis can be more peaked than the spectrum 
on the beamline axis.  Also, at these small angles 
the peak energy itself is reduced.  The narrowest neutrino energy spectrum 
occurs when the far detector is placed at an angle corresponding to
$90^\circ$ in the pion center of mass.  In this configuration, the
$\nu_e$ flux comes from the three-body decays of the muon, so the
intrinsic $\nu_e$ flux at lower energies does not increase at higher angles 
like the $\nu_\mu$ flux does.  Also, the neutral current background is
always a steeply falling function of visible energy because the
outgoing neutrino always takes some fraction of the incoming
neutrino's energy.
 
With this ``off-axis'' strategy, the \nova\ and T2K experiments still expect 
there to be some background events after all the analysis cuts are made, even 
in the absence of $\nu_\mu \to \nu_e$ oscillations.  
The measurement of the $\nu_\mu \to\nu_e$ probability requires 
knowing the level of the remaining background, and the cross section
and detection efficiencies for $\nu_e$ interactions.  
 
\subsection{Quantifying the effects due to cross section uncertainties} 
 
In order to understand why precise cross section
measurements are needed 
for a $\nu_e$ appearance experiment, it is helpful to revisit how
experiments will measure the $\nu_\mu \to \nu_e$ oscillation
probability.  The number of events in the far
detector can be described as

\begin{equation} 
N_{\textstyle far}  =  \Phi_\mu P(\nu_\mu \to \nu_e) \sigma_e
\epsilon_e M_{\textstyle far} + B_{far}  
\label{eqn:prob1}
\end{equation}
where $\Phi_\mu$ is the muon neutrino flux at the far detector, $P$ is
the oscillation probability, $\sigma_e$
and $\epsilon_e$ are the electron neutrino cross section and
efficiency, respectively, and $M_{far}$ is the far detector mass.  The
background at the far detector, $B_{far}$, can be expressed as

\begin{equation}
B_{far} = \Sigma_{i=e,\mu} \Phi_i P(\nu_i \to \nu_i)\sigma_i\epsilon_i M_{far} 
\label{eqn:prob2} 
\end{equation}       

The notation is the same as equation \ref{eqn:prob1}, but
$\epsilon_i$ is the efficiency for a neutrino of type $i$ to be
misreconstructed as an electron neutrino.  Backgrounds come from
both muon and electron neutrinos, and from several different
neutrino interaction channels.  Both equation
\ref{eqn:prob1} and \ref{eqn:prob2} must be summed over 
those channels (quasi-elastic, resonance, etc.), 
as well as integrated over neutrino energy.

The error on the oscillation probability, in this simplified notation,
is expressed as

\begin{eqnarray}
\left( \frac{\delta P}{P} \right) ^2 & = &  
\frac{N_{\textstyle far}}{(\Phi_\mu
  \sigma_e\epsilon_e M_{\textstyle far})^2} +   \nonumber \\
& & \frac{(\delta B_{\textstyle far})^2}{(\Phi_\mu
  \sigma_e\epsilon_e M_{\textstyle far})^2} +
 \frac{(N_{\textstyle far}-B_{\textstyle far})}
{  \Phi_\mu \sigma_e\epsilon_e M_{\textstyle far} } 
\left[ (\frac{d\Phi_\mu}{\Phi_\mu})^2
    + 
(\frac{\delta \sigma_e}{\sigma_e})^2 + 
(\frac{\delta \epsilon_e}{\epsilon_e})^2 \right] 
\label{eqn:proberr1} 
\end{eqnarray}

The first term comes from the statistical error on the number of events 
at the far detector.  
The second and third terms in equation \ref{eqn:proberr1} suggest two regimes: in
the case where the number of events in the far detector is comparable
to the background prediction, the error on the probability is
dominated by statistics and the uncertainty on the
background.  In the
other regime, where the number of events is dominated by the signal
events, the uncertainty on the probability is a combination of 
the statistics and
the uncertainties on the signal channel cross sections.  

Two of the three experiments described earlier in the $\nu_\mu$ disappearance
section are in fact optimized for $\nu_e$ appearance.  
Recall that T2K will use a 0.7~GeV narrow band beam and a water Cerenkov detector, 
and \nova\ will use a 2~GeV narrow band beam and a scintillator-based calorimeter.  
It is extremely important that these measurements be made at more than one 
baseline and neutrino energy, in order to be able to probe not only the 
mixing angles, but also the neutrino mass hierarchy.  In particular, only by running
at a few GeV will one be able to use matter 
effects in the earth to determine whether neutrinos follow the same mass hierarchy 
as the charged fermions.  Therefore, it is not enough to simply reduce cross 
section uncertainties below 1GeV where the cross section is predominantly 
quasi-elastic and resonance production.  To get to the mass hierarchy we will need to 
understand neutrino interactions well above a few GeV, which means also understanding 
coherent and deep inelastic scattering processes.  

\subsection{Cross Section Uncertainties with a Near Detector} 

     Both \nova\ and T2K plan to make far detector event predictions based on 
measurements made in near detectors.  For the case of \nova\ 
the near detector is 
planned to be of a very similar design to the far detector, 
and can be placed in a wide range of angles with respect to the 
NuMI beamline.  By making the near 
detector similar, \nova\ hopes to minimize 
uncertainties in the detector response and 
efficiency.  However, because the near detector 
will be as coarse as the far, it is 
not optimized for cross section measurements.   For the T2K near 
detector suite some 280~m from the proton target, 
the plan is to have one near detector on axis to measure the
spectrum and transverse distribution, 
and at least one other near detector 
that is off-axis which will be focused on cross section
measurements.  There are longer term 
plans to build a water Cerenkov detector at 2~km 
from the proton target, but even then 
the detector is not modular and as such 
the efficiencies are not expected to be identical
between the near and far detectors.  

     To see how any uncertainties (cross section, detector 
acceptance, or flux) will arise in 
the far detector prediction based on the near detector data, 
it is useful to think about how
the event samples are likely to change between near and far.  
At a near detector, the flux of muon neutrinos will have a 
very strong peak at a particular energy, while at the 
far detector that peak will have oscillated mostly to 
$\nu_\tau$'s.  At these energies, $\nu_\tau$'s will not produce
charged current events, only neutral current events.  
The neutral current event samples
are likely to be similar from near to far, provided the
near detector is at a similar off-axis angle.  The electron
neutrino events at the peak are primarily from muon decays
in the beamline, which occur on average substantially farther
downstream than the pion decays.  Therefore, the extrapolation 
from the near to far detector tends to be different for all 
three event samples.  If one cannot predict for the near event 
sample how many background events belong to each category
(due to any of the above uncertainties), the far detector  
extrapolation can be wrong.  

As a quantitative example of how cross section uncertainties 
would not completely 
cancel between near and far detectors, 
a study was done using a simulation for an early design\cite{rpc}  
of the \nova\ detector. Although the final design of the \nova\ detector 
will be different, the fundamental arguments will still be true:  
there will be a mix of contributing
cross sections at the far detector that by definition cannot be the same
mix as that at the near detector.  

The
signal and background statistics for the nominal 5 year run are given
in table \ref{tab:nova1}.  
Also given in table \ref{tab:nova1} are the fractions that each
neutrino interaction process contributes to the events of that type
that pass all cuts, as well as the cross section uncertainty on that
process, as tabulated in reference~\cite{sam}.
Without a near detector, the total error on the background prediction 
from cross section uncertainties,
for the case that there are no $\nu_\mu$ oscillations, is 16\% , which
is equivalent to the statistical error for that case.  For the
case of mixing at the level indicated in the table, the statistical
error on the probability would be 8\% , while
the errors from cross section uncertainties alone would be 31\% .

\begin{table}[th] 
\begin{center} 
\begin{tabular}{|l|c|c|c|c|c|} 
\hline
 &  & QE & RES & COH & DIS \\ \hline
  &  & \multicolumn{4}{|c|}{Cross Section Uncertainty}\\ \hline
 &   & 20\% & 40\% & 100\% & 20\% \\ \hline
  &  & \multicolumn{4}{|c|}{Composition after all cuts }\\
 Process &Statistics & \multicolumn{4}{|c|}{in far detector }\\\hline 
Signal $\nu_e$ & 175 ($\sin^2 2\theta_{13}=0.1)$ & 55\% & 35\% & n/I & 10\% \\ \hline
NC & 15.4 & 0 & 50\% & 20\% & 30\% \\ \hline
$\nu_\mu CC $ & 3.6 & 0 & 65\% & n/I & 35\% \\ \hline
Beam $\nu_e$ & 19.1 & 50\% & 40\% & n/I & 10\% \\ \hline
\end{tabular}
\end{center} 
\caption{List of the signal and background processes than can
contribute events in the \nova\ far detector, for a 50~kton detector
located 12~km from the NuMI axis, 820~km from Fermilab, assuming a
$\Delta m_{23}^2$ of $2.5\times 10^{-3}eV^2$.  Also given are the current 
cross section uncertainties on those processes. ``n/I'' indicates that 
the charged current coherent process was not included, since it is 
expected to be small compared to other charged current processes.}
\label{tab:nova1}
\end{table}

\begin{figure}[tp]
\begin{minipage}{.5\textwidth}
\epsfxsize=\textwidth
\epsfbox{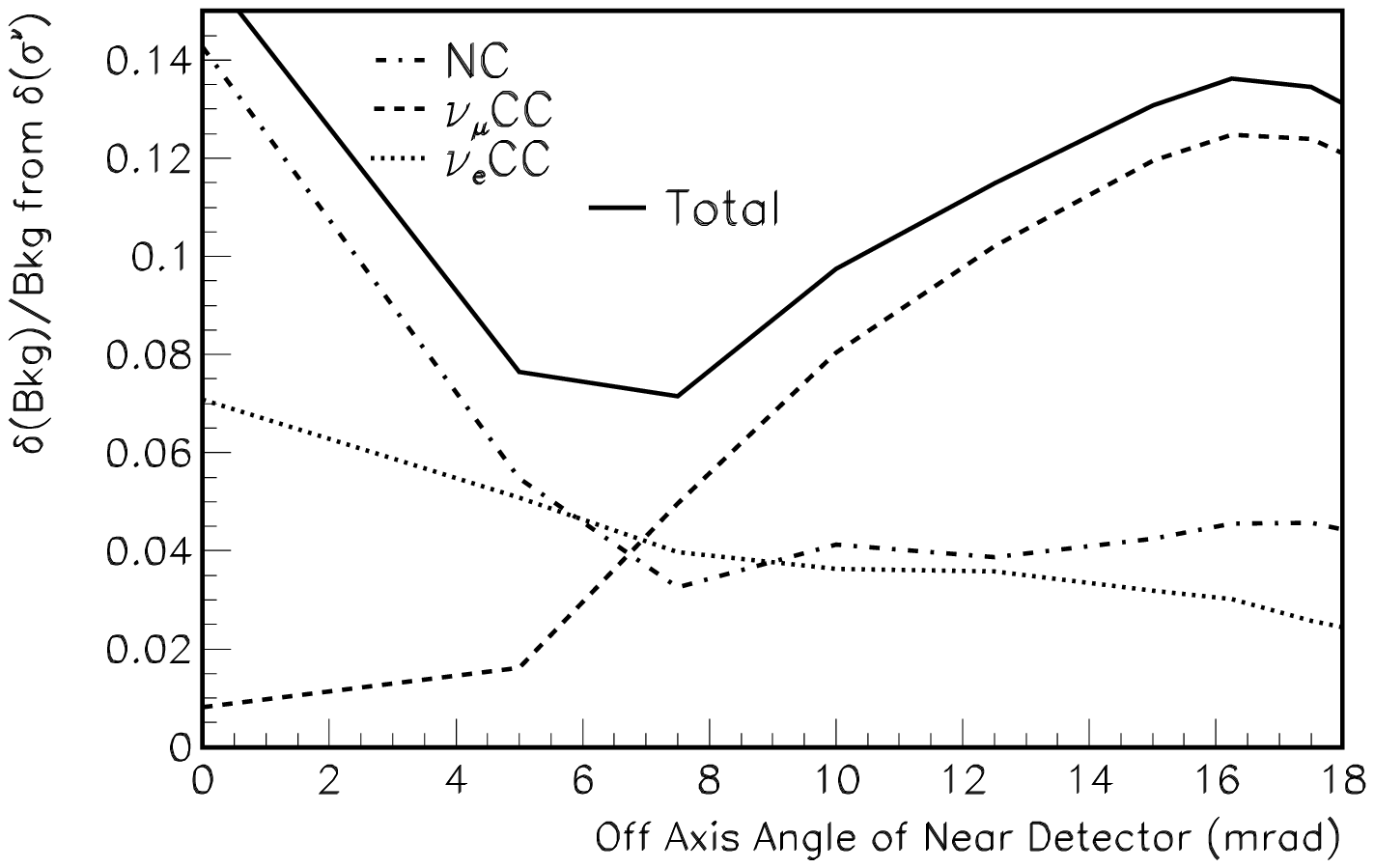}
\end{minipage} 
\begin{minipage}{.5\textwidth}
\epsfxsize=\textwidth
\epsfbox{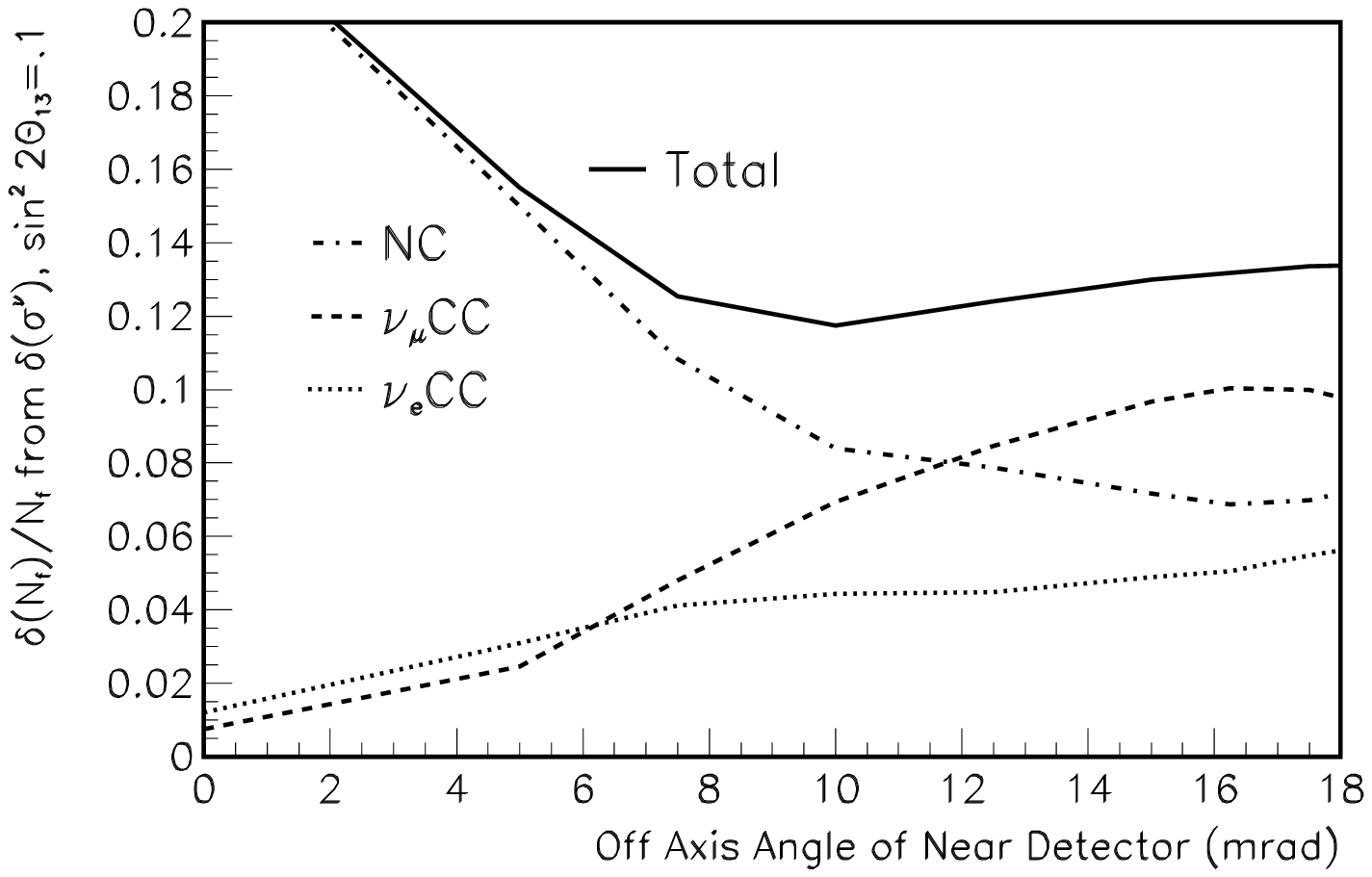}
\end{minipage} 
\caption{The fractional error in the event rates at the far detector 
from uncertainties in each process (Quasi-elastic, resonance, deep inelastic
scattering, and neutral current coherent $\pi^0$ production), 
added in quadrature for each source (neutral current, 
$\nu_\mu$ charged current, beam $\nu_e$), plotted
as a function of the angle between the near detector 
and the beamline axis, for (left) background-dominated
experiment and (right) signal-dominated experiment.}
\label{fig:toterr}
\end{figure}

Figure \ref{fig:toterr} shows the fractional error on the 
far detector prediction 
as a function of the angle between the beamline and the 
near detector, 
for two different extremes:  the left plot shows 
the case where the $\nu_\mu \to \nu_e$ probability 
is zero (corresponding to the background-limited experiment), and 
the right plot shows the case where 
the probability is at about 5\% (or $\sin^2 2\theta_{13}=0.1$, 
corresponding to the signal-dominated experiment).
For low angles the error due to the high $y$ 
$\nu_\mu$ charged current uncertainties is smallest.
For high angles the errors due to neutral current uncertainties
and low $y$ $\nu_e$ charged current uncertainties are the smallest.

The errors for each of the
three background contributions are shown, where the errors due to 
quasi-elastic, resonance, DIS, and coherent cross section
uncertainties are added in quadrature.  
In the case of the background-dominated experiment, the cross section 
errors alone are comparable no less than half the expected statistical 
error of about 15\% .
For the signal-dominated experiment, the cross section at best a factor of 
two worse than the expected statistical error of 7\% .  

\subsection{Future Measurements of Low Energy Cross Sections} 

Given the low statistics, discrepant data, 
and limited reach in target nuclei for 
charged and neutral current cross section measurements, there is 
clearly much work to be done.  Section \ref{section:xsecerr} described 
the cross section uncertainties for quasi-elastic and resonance 
charged current processes, and described how \minerva\ could provide
an accurate quasi-elastic cross section measurement.  For 
$\nu_e$ appearance measurements the charged current cross sections 
are important in case of a large signal. Regardless of 
signal size, however, the neutral current cross sections are 
important since they are very 
poorly known now. 
In some cases the best strategy will be to measure the charged 
current analog as a function of neutrino energy, 
and depend on theory combined with an average neutral
current measurement to predict the neutral current cross section
as a function of neutrino energy.  Recent neutral current
measurements have been normalized to different charged current channels:  
for example, the 
ratio of single $\pi^0$ production in 
neutral currents to the total $\nu_\mu$ charged
current cross section has been measured 
to about 11\% by the K2K collaboration \cite{k2kpi0}.  

With an appropriate design that would include both fine-grained fully active 
target surrounded by electromagnetic and hadronic 
calorimetry, the uncertainties on these cross sections
could be improved by factors of 5 or more.  As an example, 
the \minerva\ experiment proposes to 
reduce the relevant cross section uncertainties for \nova\ 
to about 5\% for all of the 
charged current and neutral current DIS processes, 
10\% for the neutral current resonance processes, 
and 20\% for the neutral current 
coherent $\pi^0$ processes \cite{minervaprop}.  
But before describing how these
measurements would be made, it is striking to see how 
much these measurements would reduce the systematic errors 
shown in figure \ref{fig:toterr}.    

\begin{figure}[tp]
\begin{minipage}{.5\textwidth}
\epsfxsize=\textwidth
\epsfbox{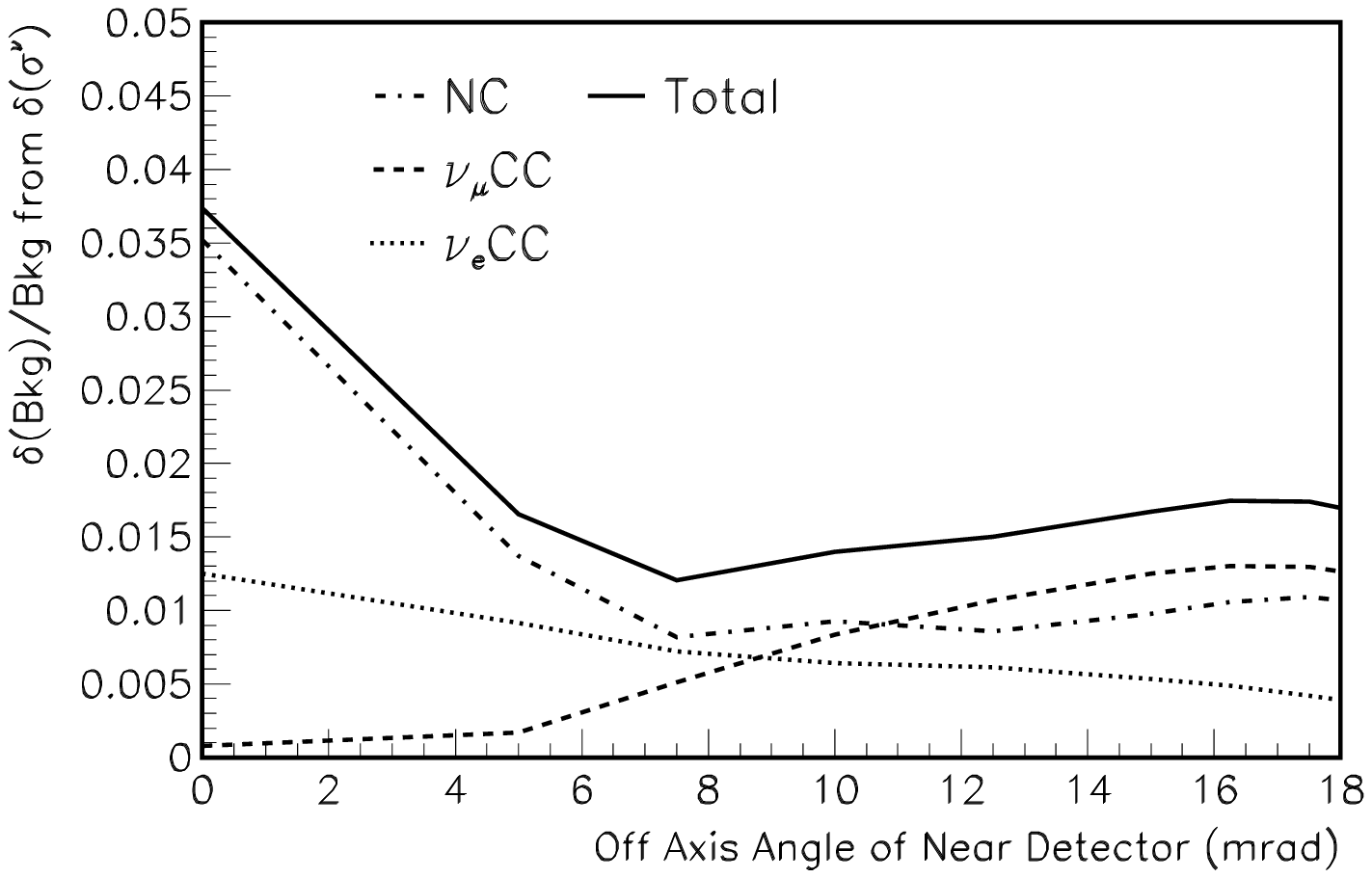}
\end{minipage}
\begin{minipage}{.5\textwidth}
\epsfxsize=\textwidth
\epsfbox{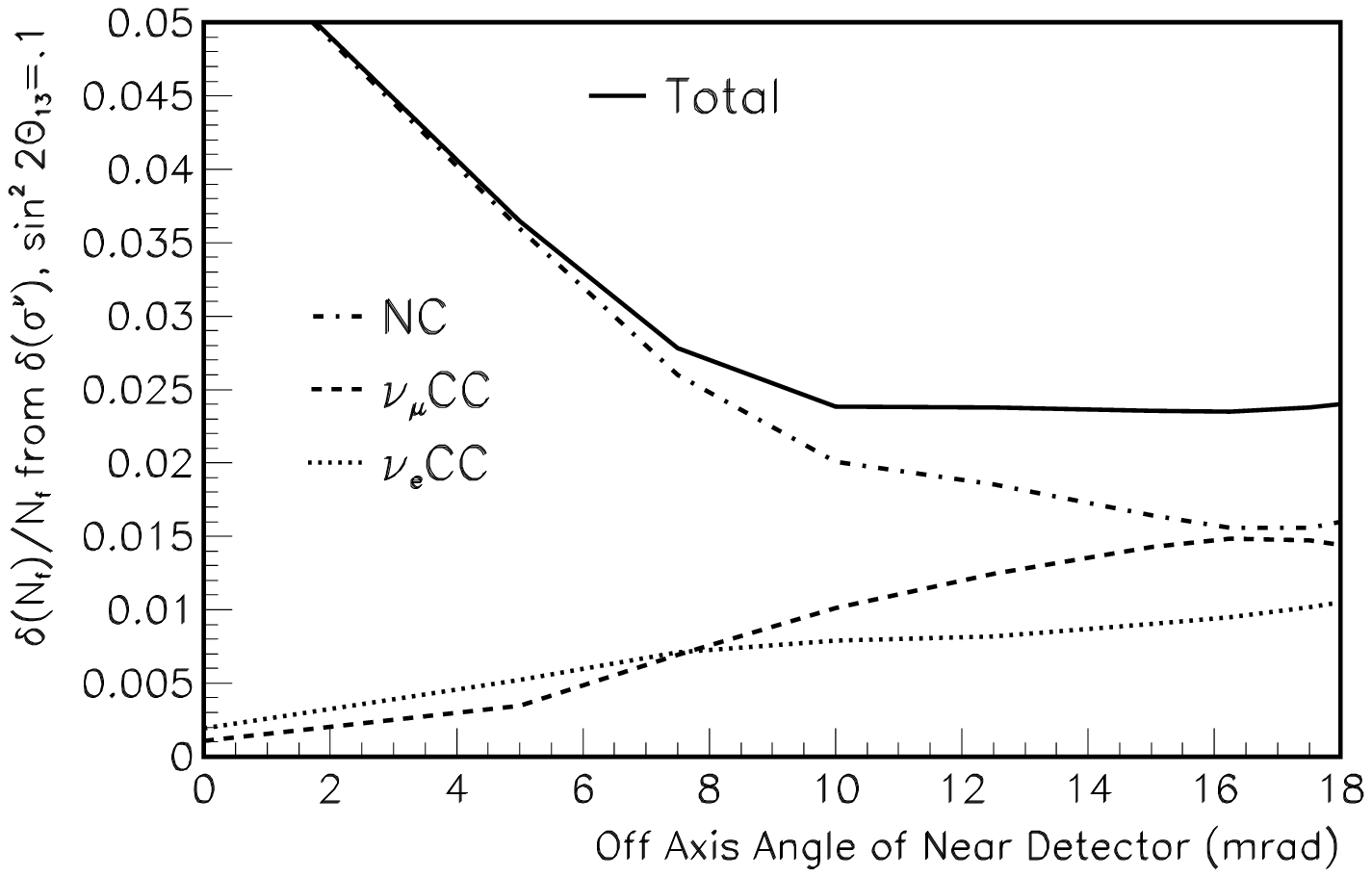}
\end{minipage}

\caption{The fractional error in the total event rate at the far detector 
from post-\minerva\ uncertainties in each process
as a function of the angle between the near detector 
and the beamline axis, for the case where the 
$\nu_\mu \to \nu_e$ probability is 0 (left) or 5\% (right).}
\label{fig:postminerva}
\end{figure}

If the uncertainties described above were achieved, 
then the systematic errors due to cross section uncertainties
would be well below the statistical errors, as shown in 
figure \ref{fig:postminerva}.  
For the background-dominated experiment (left), the systematic error would be 
about a factor of ten less than the statistical error, 
and for the signal-dominated
experiment (right) the systematic error would be a factor of three 
below the statistical error.  

The remainder of this article describes strategies for isolating the 
resonant and coherent cross sections in the \minerva\ detector, and 
the expected statistical precision in a four year run. 

\subsubsection{Resonance Cross Sections} 

     Resonance production in neutrino scattering is extremely 
important for future long baseline neutrino oscillation experiments, 
but its 
cross section is only known at about the 40\% level for the charged current 
process~\cite{ccsinglepi0} at 2~GeV, and much worse for the 
neutral current process~\cite{ncsinglepi0}.  

     Resonance production can be studied in detail with a fine-grained  
experiment with good vertexing abilities and a low
threshold for seeing pions.  By requiring an outgoing muon, pion, and proton, 
\minerva\ expects to fully reconstruct a large fraction of the 
$2\times 10^5$ charged current resonance events that will occur in the detector, 
which
would enable not only a precise cross section measurement as a function of 
energy, but also enough statistics to measure the $W^2$ distributions.
With good neutral and charged pion identification the individual states containing
both charged and neutral pions can be clearly seen, which in turn are important
for $\nu_\mu$ disappearance and $\nu_e$ appearance, respectively.  

By measuring 
charged and neutral current resonance production and combining this with the 
energy information from the charged current resonance production, models
that relate charged to neutral currents will be tested, and precise predictions
for the neutral current processes will become available.  

\subsubsection{Coherent Cross sections} 

\begin{figure}[tp]
\begin{minipage}{.5\textwidth}
\epsfxsize=\textwidth
\epsfbox{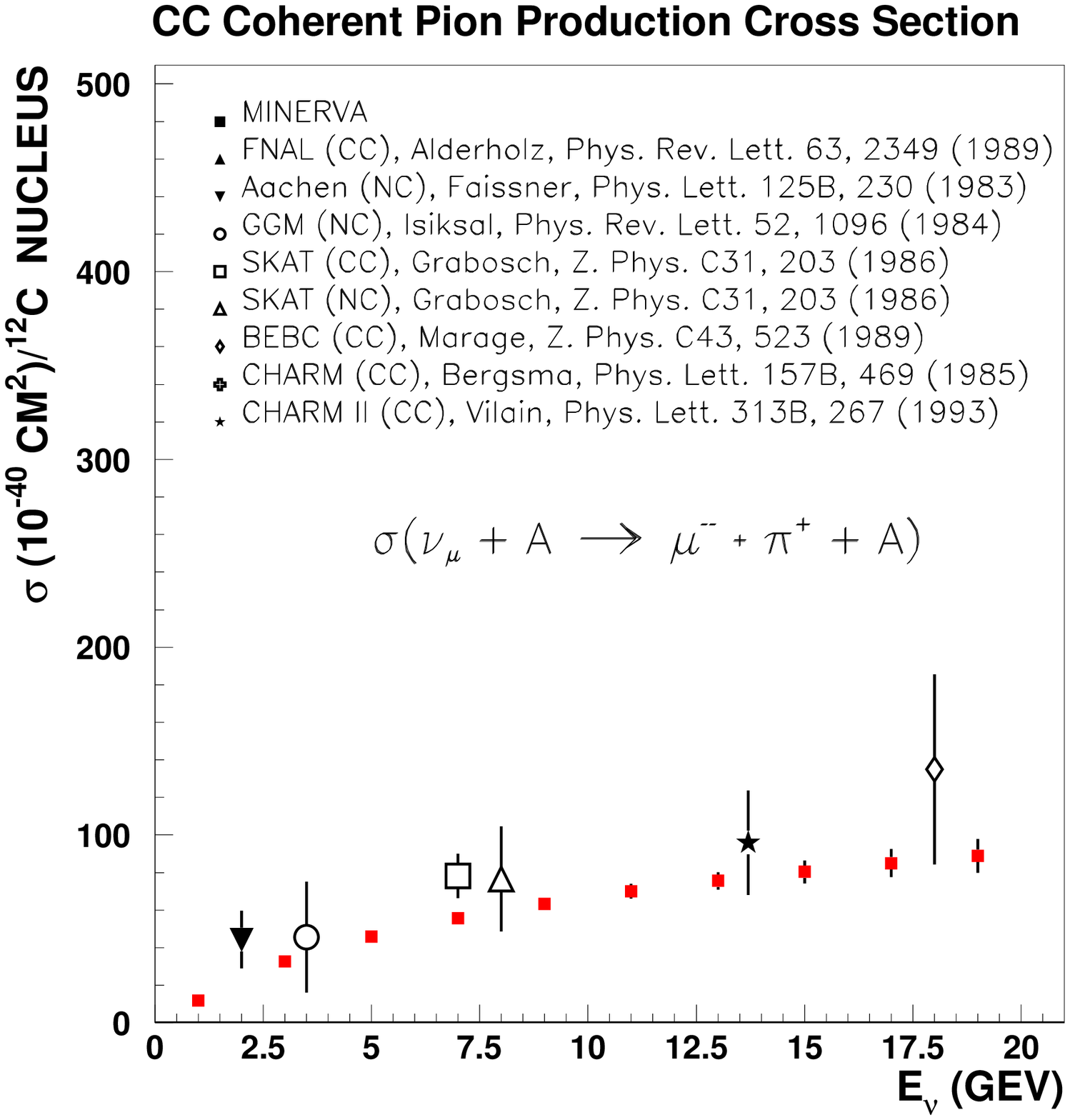}
\end{minipage} 
\begin{minipage}{.5\textwidth}
\epsfxsize=\textwidth
\epsfbox{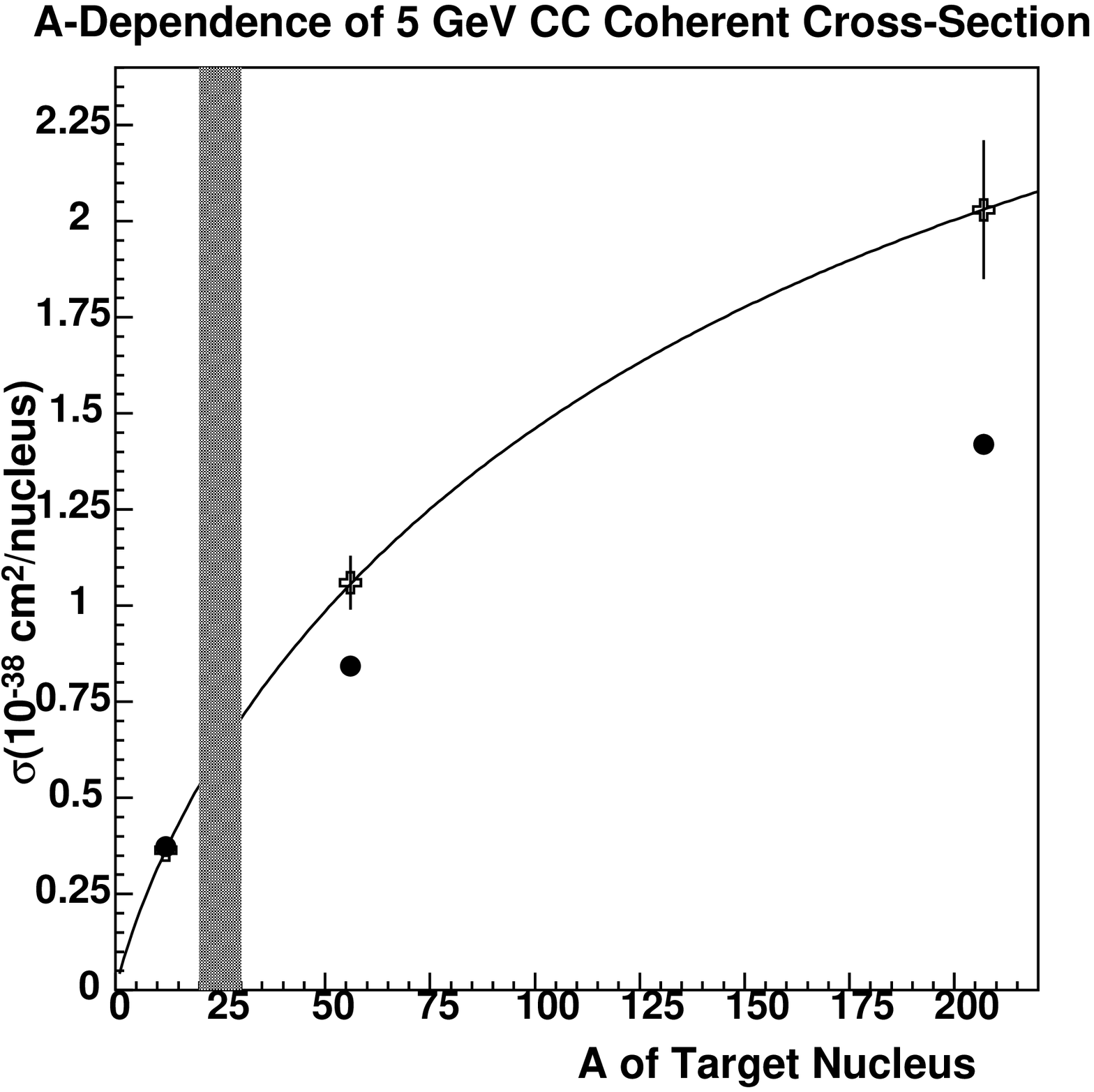}
\end{minipage} 
\caption{Expected \minerva\ statistical 
sensitivity for the charged current 
coherent cross section energy (left) and $A$ (right) 
dependence measurements, for a 4 year parasitic 
MINOS run, taking into account detector
acceptance.  }
\label{fig:mincoh}
\end{figure}

     The process by which a neutrino interacts with a nucleus coherently and 
produces only a neutral pion (in the neutral current process) or a muon and 
a charged pion (in the charged current process) is perhaps the process the 
most poorly measured yet still seen.  A handful of measurements exist at 
the few sigma level in both the neutral 
(\cite{nccoh}) and charged (\cite{cccoh}) 
current channels, as shown in figure \ref{fig:mincoh}(left).  
Although the cross section for this process is low, its high uncertainty
and the high probability that coherent events pass $\nu_e$ analysis cuts
means that 
this channel will contribute a significant uncertainty in the neutral current 
background.  Furthermore, because it is an 
interaction that does not break up the nucleus, the nuclear effects on 
the cross section are important.   

Coherent charged current events can be identified by looking at the 
energy loss of the two tracks and requiring it to be consistent with 
the presence of a muon and a pion, and nothing else.  The background 
would come from incoherent processes 
where other particles (for example a proton) were lost.  
Coherent neutral current events would be identified
by looking for two electromagnetic showers which reconstruct to the pion 
invariant mass.  Backgrounds here would 
come again from incoherent processes, and 
are expected to be larger because 
several processes produce at least one neutral pion.  
The neutral current coherent sample can 
be separated statistically by looking at the 
distribution of the reconstructed angle of the 
neutral pion with respect to the neutrino direction
and subtracting the background under 
the forward scattering peak.   

The \minerva\ experiment running in the NuMI beamline would collect 
over a thousand charged and neutral current coherent events 
in a 3-ton fiducial volume per year, resulting in a precise measurement
as a function of neutrino energy for the charged current process.  Figure
\ref{fig:mincoh} shows both the 
energy (left) and atomic number (right) dependence
that could be measured by \minerva\, in the charged current channel 
along with the current set of measurements.  By 
using theory and the high statistics neutral current data 
one could obtain at least a factor of 
five improvement in the precision on the 
neutral current coherent background prediction.

\section{Conclusions} 

It is clear from even these preliminary studies that dedicated
neutrino scattering experiments such as \minerva\ will play a
very important role in helping the current and future precision
oscillation experiments reach their ultimate sensitivity.  In order to
get the most precise values of $\Delta m^2_{23}$ (which eventually is
used to extract mixing angles and the CP-violating phase) this field
must better understand and quantify the processes that occur between
the interaction of 
an incoming neutrino and the measurement of the 
outgoing particles in the 
detectors.  Although the issues are different depending on whether
those detectors are water Cerenkov or calorimetric devices, in both
cases more information is needed.  
Extracting the mixing parameters
such as $\theta_{13}$ and ultimately the neutrino mass hierarchy and
CP violation requires much better understanding of resonant 
cross sections.  Even setting limits on these parameters will
require better measurements of neutral current processes.  Precise
measurements of nuclear effects and exclusive cross sections will lay
an important foundation for a field that is in the middle of making
order of magnitude leaps in both statistics and sensivitity.


\begin{thebibliography}{00}
\bibitem{rpc}  The simulation assumed the active material was resistive plate
chambers and the absorber was particle board (hydrocarbons).
\end{thebibliography}
\end{document}